\documentclass[aps,prb,preprint,superscriptaddress,showpacs]{revtex4}
\usepackage{graphicx}
\usepackage{latexsym}

\begin{document}


\title{Origin of the anomalous Slater-Pauling curve in CoMn alloy clusters}


\author{M. Pereiro}
\email[Corresponding author. ]{manuel.pereiro.lopez@usc.es}
\affiliation{Departamento de F\'{\i}sica Aplicada and Instituto de Investigaci\'ons Tecnol\'oxicas, Universidade de Santiago de Compostela, Santiago de Compostela E-15782, Spain}
\author{D. Baldomir}
\affiliation{Departamento de F\'{\i}sica Aplicada and Instituto de Investigaci\'ons Tecnol\'oxicas, Universidade de Santiago de Compostela, Santiago de Compostela E-15782, Spain}
\author{J. E. Arias}
\affiliation{Instituto de Investigaci\'ons Tecnol\'oxicas, Universidade de Santiago de Compostela, Santiago de Compostela E-15782, Spain}


\date{\today}

\begin{abstract}
Surprising enhancement of the magnetic moments recently observed in dilute CoMn alloy clusters is explained using
{\it ab initio} electronic structure calculations. The calculated magnetic moments generally agree with
the reported experimental data. An equation for calculating the magnetic moments of the CoMn alloy clusters has been
derived to correct the deviations predicted by the rigid-band model and the virtual bound states approximation. 
A new strategy is proposed to obtain the ground-state structures of the CoMn
clusters and it was also put to the test of the experiment. 
\end{abstract}


\maketitle
\section{Introduction}
The exploration of bimetallic transition metal (TM) clusters is emerging as a promise
field of research because of the new opportunities they offer for developing magnetic recording devices and cluster-assembled
materials with new functions for medical applications \cite{bansmann,binns}. Although the electronic and magnetic properties of bare TM
clusters have been actively studied for several years, 
however less attention has been paid to their alloys because they represent both an experimental and theoretical challenge.
Thus, for experimentalists is very difficult to control the stoichiometry of the alloy clusters using chemical methods, and for 
theoreticians, the determination of the ground-state geometries becomes a very difficult task, as commented later 
in the main text. 
The recent observation of the average magnetic moment enhancement in
dilute CoMn alloy clusters \cite{yin} that contradicts the bulk behavior 
has introduced an entirely unexpected dimension on the subfield
of magnetic alloy clusters and it could pave the way for future possible applications like, for example, in biomedicine 
as magnetic sensors. 
Although the search for the origin of the aforementioned observation has 
actively stimulated the research on CoMn alloy clusters \cite{yin,rollmann,ganguly,shen}, however the answer still remains elusive. A possible explanation
is based on the assumption that a virtual bound state (VBS) is formed below the highest occupied molecular orbital (HOMO) level and near the Mn site \cite{yin}. 
A VBS can be defined in the potential scattering model as a resonant scattering near impurity atoms in 
the host which induces a narrow peak 
in the conduction band density of states (DOS). Originally, the VBS model was developed by Friedel \cite{friedel} 
to explain many of the
physical properties of bulk magnetic alloys containing dilute magnetic species. It represents an improvement over 
the rigid-band (RB) model which is based on the assumption that the $s$ and $d$ bands are rigid in shape
as atomic number of the alloy changes \cite{slater}. In this article, our calculations demonstrate that the origin 
of the anomalous behavior of the Slater-Pauling (SP) curve
of CoMn alloy clusters does not require the formation of a VBS as suggested in Ref.~\onlinecite{yin}, but is explained 
directly in terms of the magnetic moment provided by the Mn atoms and the ``spin-flipping'' of the electrons belonging
to the Co-Mn bonding. Thus, with the aim of studying the magnetic properties 
of Co-Mn alloy clusters and elucidating the anomalous behavior of their SP curve,
we have performed spin-unrestricted density functional theory-based (DFT) calculations 
as implemented in the {\sc demon-ks3p5} program package~\cite{salahub}. 

\section{Method and structural properties}
The electronic system consisting in $\Omega$ nuclei and $\eta$ electrons is assumed to be described by the
next Hamiltonian, which expressed in second quantization and in the Born-Oppenheimer nonrelativistic approximation reads as
\begin{eqnarray}
  \label{eq:1}
\lefteqn{\hat{H}=\sum_{ij}^{\eta} \sum_{\nu\mu} c^*_{\nu i}c_{\mu j}\left(\left< \nu\right| 
	-\frac{1}{2}\triangle\left|\mu\right>
	-\sum_k^{\Omega}Z_k\left<\nu\right|\left|\mu\right>\right) \hat{a}_i^\dagger \hat{a}_j}\hspace{40pt} \nonumber\\
 &&+\frac{1}{2} \sum_{ijkl}^{\eta}\sum_{\nu\mu\alpha\beta} c_{\nu i}^* c_{\mu j}^* c_{\alpha k} c_{\beta l} \left<\nu\mu\right|\left|\alpha\beta\right> \hat{a}_i^\dagger\hat{a}_k^\dagger\hat{a}_l\hat{a}_j  \nonumber \\ 
 && + \hat{V}_{xc}^{\mbox{\tiny GGA}}\left(\rho^\sigma,\nabla \rho^\sigma\right)  
\end{eqnarray}
where the Kohn-Sham orbitals ($\left|i\right>$) were expanded into atomic orbitals 
$\left|i\right>=\sum_{\mu} c_{\mu i}\left|\mu\right>$  in the Linear Combination of Gaussian 
Type-Orbitals ansatz. The orbital basis sets of contraction pattern (2111/211*/311+) and (2211/311/411) 
were used in conjunction
with the corresponding (5,3,3;5,3,3) and (3,4;3,4) auxiliary basis sets for Co and Mn, respectively~\cite{huzinaga}.
Likewise, {\it ad hoc} model core potentials with contraction pattern (4:7,4) and (5:7,4) have been used for describing the inner
electrons of Co and Mn atoms, respectively~\cite{pereiro}.
The kinetic  and nuclear attraction  energy 
of the electrons in an environment of $Z_k$ nuclear charges are described by the one-electron term. The $||$ symbol
represents the $1/|\mathbf{r}-\mathbf{k}|$ operator, where $\mathbf{r}$ and $\mathbf{k}$ are the electron and 
nuclear position vectors, respectively.
The two-electron operator represents the Coulomb repulsion  energy of the electrons. In this case,  
the symbol $||$ represents the $1/|\mathbf{r}_1-\mathbf{r}_2|$ operator for the electrons with 
coordinates $\mathbf{r}_1$ and $\mathbf{r}_2$. The last term in Eq.~(\ref{eq:1}) is the exchange-correlation 
energy and we have used here the form proposed in Ref.~\onlinecite{perdew} which is a 
function of the spin-dependent electron density ($\rho^\sigma, \sigma=\uparrow,\downarrow$) and its gradient 
in the generalized gradient approximation (GGA). We have adopted the former GGA functional 
in our DFT calculations because it was reported in 
Ref.~\onlinecite{pereiro} that it represents
a dramatical improvement in the calculated binding energies of Co clusters.
The total energy of the system is calculated adding
the nuclear repulsion energy ($\sum_{a>b}^\Omega (Z_a Z_b)/|\mathbf{r}_a-\mathbf{r}_b|$) to the 
electronic contribution. A wide set of spin multiplicities ranging from 1 up to a maximum of 61, depending on
the selected cluster, was checked to
ensure that the lowest-energy electronic and magnetic configuration is reached. More information about the computational
details can be found elsewhere~\cite{pereiro1}.  Hereafter, all calculated results refer only to 
the Co-Mn alloy 20 atom clusters. We have also performed electronic structure calculations of some guessed
geometries of the Co-Mn 
alloy clusters with $\Omega=25$ and 30 in a range of Mn concentration less than 0.3, but 
the obtained results slightly differ from that of the $\Omega=20$ case, i.e., we observed that the calculated
magnetic moments of clusters with $\Omega=25$ and 30 increase with the impurity concentration as $\Omega=20$ clusters do.
Accordingly, the conclusions drawn from the Co-Mn clusters with $\Omega=20$ can be extended to clusters with greater size.

\begin{figure}
  \includegraphics[width=16.5cm,angle=0]{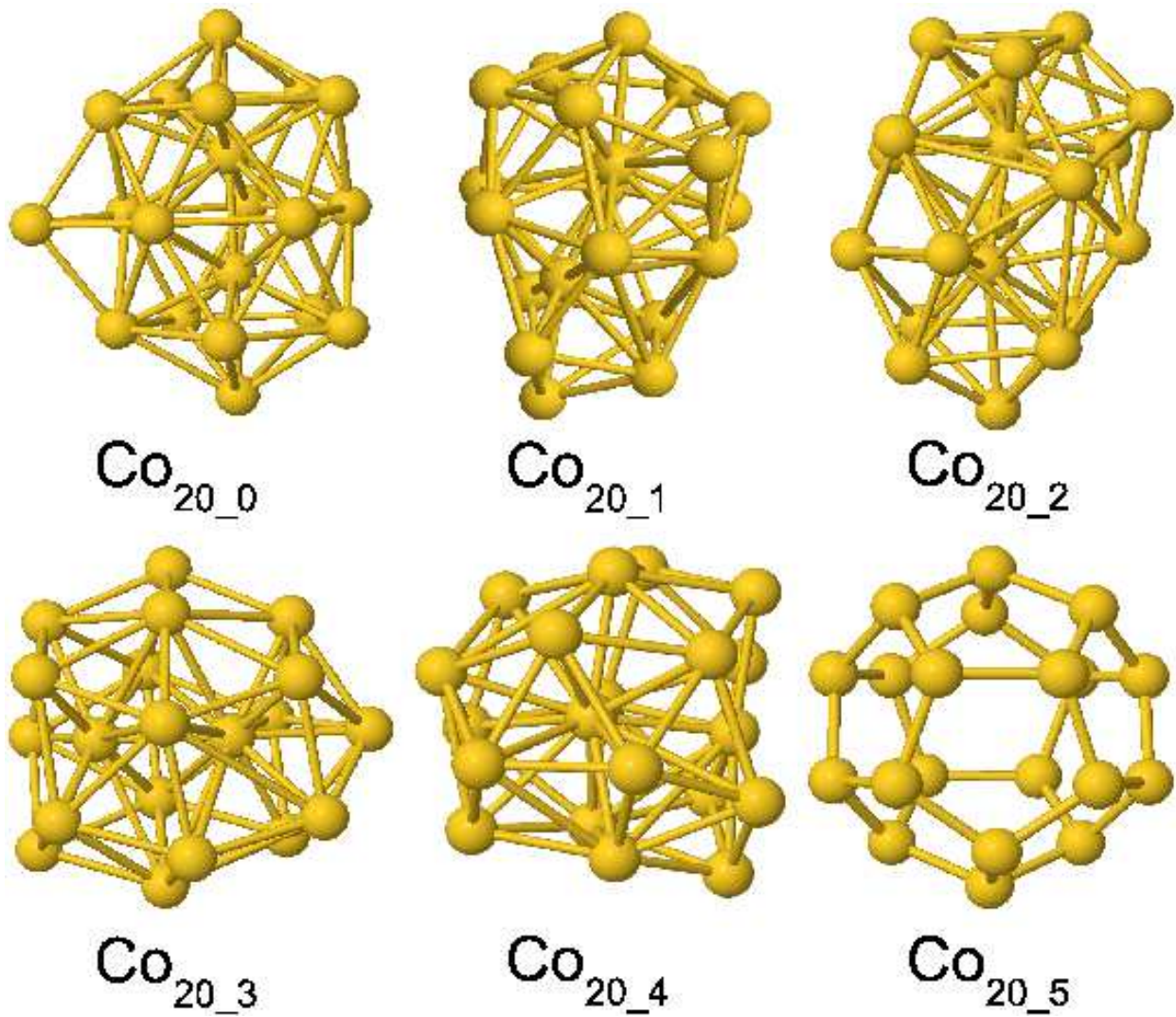}
  \caption{\label{fig1}(Color online) Illustration of the guessed structures of the Co$_{20}$ cluster. The structures
  are ordered from left to right and top to bottom by increased relative energy. The notation Co$_{20\_m}$ stands for 
  the $m$th energetic isomer.}
\end{figure}

\begin{figure}
  \includegraphics[width=16.5cm,angle=0]{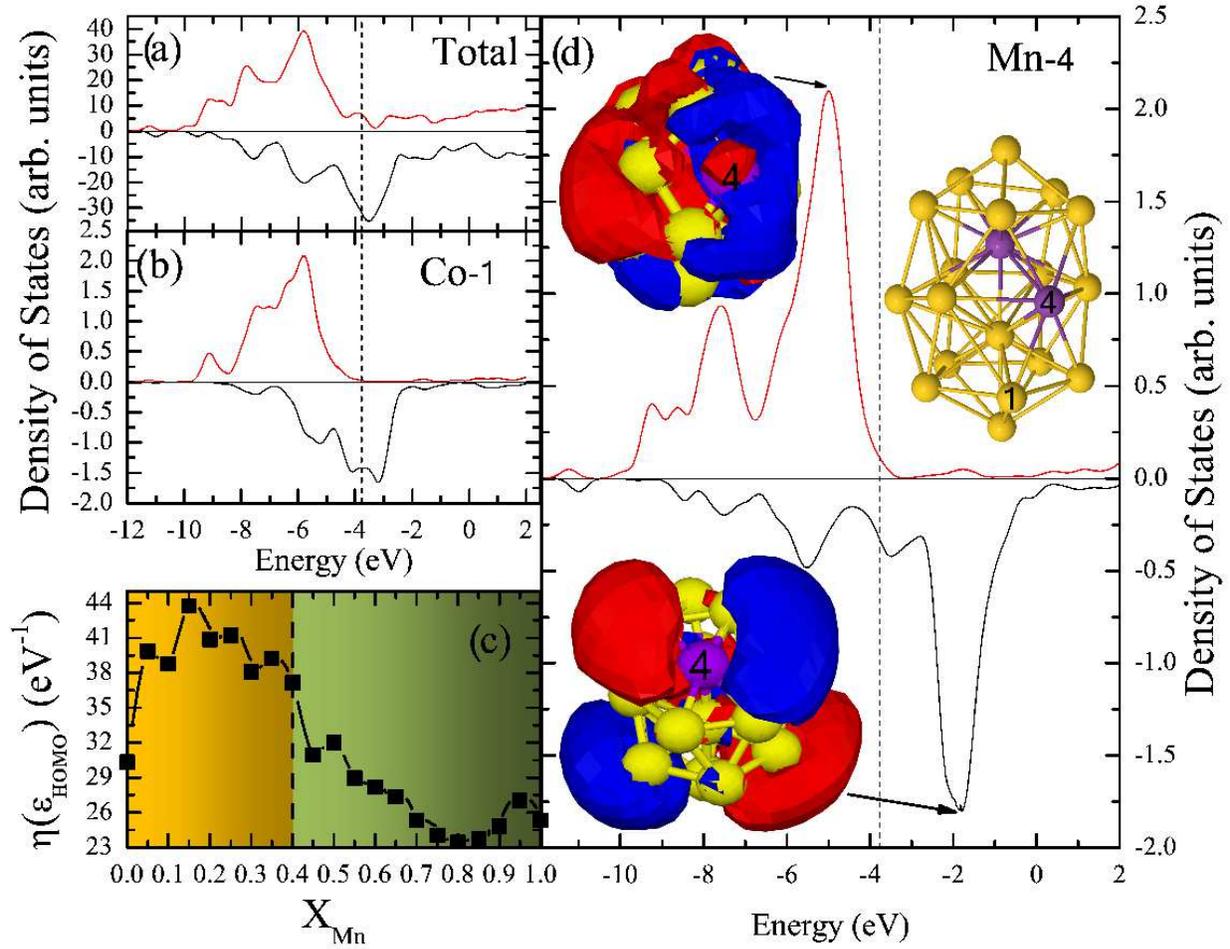}
  \caption{\label{fig2}(Color online) Plot of the spin-polarized DOS for (a) Co$_{18}$Mn$_2$, (b) Co-1 and (d) Mn-4, where the
  labels 1 and 4 represent the site number of Co and Mn atoms, respectively. In (b) 
and (d), we have only plotted the contribution of the $d$ orbitals because $s$ and $p$ 
orbitals are of less importance. The upper-half
part of the figures is for spin-up electrons while the lower-half is for the spin-down electrons. The atoms 
of (b) and (d) are labeled in the geometric structure. We have also plotted the shape of the delocalized molecular orbitals
for the higher-lying occupied and lower-lying unoccupied levels of Co$_{18}$Mn$_2$ cluster with the aim of showing the 
bonding and antibonding character of the orbitals, respectively. The dotted vertical lines 
represent the HOMO level. In (c), DOS at the HOMO level is plotted against the Mn concentration.
}
\end{figure}

The search for the global minima of the Co-Mn alloys 20 atom clusters was planned as a multistage strategy combining an unbiased
search method i.e.,  a basin-hopping~\cite{wales} algorithm, in conjuction with
a molecular mechanics method~\cite{hyperchem}.
In the reoptimization procedure of the clusters, we have made use of the Polak-Ribi\`{e}re 
algorithm~\cite{fletcher} without any symmetry constraint and the root-mean-square gradient was set to $10^{-4}$ Kcal/(mol \AA). 
In a first stage of the calculation method, 
the initial guessed structures of the Co$_{20}$ cluster were taken from three different sources, namely, 
the structures were provided
by the {\sl GMIN} code~\cite{wales} which uses the basin-hoping algorithm, the existing databases~\cite{database}, and proposed by us. After that,
they were reoptimized with the {\sl HyperChem} code~\cite{hyperchem} with the intention to obtain the ideal candidate 
to the lowest-energy structure of the Co$_{20}$ cluster. The converged guessed structures are illustrated in Fig.~\ref{fig1}
and the structural and energetic parameters are reported in Table~\ref{table1}. The ground-state structure (Co$_{20\_0}$) is
a capped double icosahedron. The cohesive energy and the number of nearest-neighbor
Co-Co bonds reveal that the most stable structure is also the most compact one although the average first-neighbor distance
is a little bit higher than the rest of structures plotted in Fig.~\ref{fig1}.  
In a second stage, the rest of guessed structures 
of the Co$_m$Mn$_n$ alloy clusters were built from the geometry of the Co$_{20}$ cluster but this time 
$m-n$ Co atoms swap positions with $n$ Mn atoms. The assumption is underpinned by three arguments: 
firstly, the ionic radii (r$_a$) of Co and Mn
atoms are pretty close i. e., 1.25~\AA~and 1.26~\AA, respectively~\cite{kittel}. Secondly, 
we have
calculated the size of the Thomas-Fermi screening~\cite{friedel} outside the atomic sphere of the Mn atoms 
($Z_{Mn}$)
\begin{equation}
  \label{eq:2}
  Z_{TF}=Z_{Mn}\left(1+ r_a \sqrt{4 \pi \eta(\epsilon_{HOMO})} \right) e^{-r_a\sqrt{4 \pi \eta(\epsilon_{HOMO})} }
\end{equation}
and it is vanishingly small because the DOS at the HOMO level ($\eta(\epsilon_{HOMO})$) provided by our DFT
calculations are large [see Fig.~\ref{fig2}(c)], so that the exponent in Eq.~(\ref{eq:2}) is 
$r_a\sqrt{4 \pi \left<\eta(\epsilon_{HOMO})\right>}\simeq247.58$, where $\left<\eta(\epsilon_{HOMO})\right>$ represents 
the average over the lowest-energy Co-Mn alloy series clusters with $\Omega=20$. Thus, 
the size effects are negligible and the addition of Mn atoms to the cluster should slightly distort their environment.
Thirdly, we have performed a geometry optimization of   
Co$_3$, Co$_2$Mn$_1$, Co$_4$, and Co$_3$Mn$_1$ clusters by means of spin-unrestricted DFT calculations~\cite{salahub} 
and the results show that the elongation of the Co-Mn average bond length with respect to the Co-Co one is less than 0.15~\AA.
In consequence, the substitution of cobalt atoms by Mn atoms should slightly distort the geometric structure of the
clusters. This fact has also been observed in Ref.~\onlinecite{yin}.

\begin{table}
  \caption{\label{table1}Average first-neighbor distance (d$_\mathrm{Co-Co}$), number of 
  nearest-neighbor Co-Co bonds ($N_\mathrm{Co-Co}$) per atom,
  relative ($\Delta E$), and cohesive (E$_c$) energies of Co$_{20}$ cluster isomers.
  The geometry notation is that of Fig.~\ref{fig1}. The point groups are determined from Ref.~[\onlinecite{colby}].}
\begin{ruledtabular}
\begin{tabular}{cccccc}
          &             &  d$_\mathrm{Co-Co}$ & $N_\mathrm{Co-Co}$/atom & $\Delta E$ & E$_c$ \\
  Cluster & Point group &          (\AA)      &                         &    (eV)    & (eV)  \\
\hline
Co$_{20\_0}$ & $C_{2v}$ & 2.46 & 3.55 & 0.00 & 1.78  \\
Co$_{20\_1}$ & $C_{1}$  & 2.39 & 3.00 & 0.26 & 1.76  \\
Co$_{20\_2}$ & $C_{1}$  & 2.38 & 3.05 & 0.35 & 1.76  \\
Co$_{20\_3}$ & $C_{1}$  & 2.39 & 2.85 & 0.95 & 1.73  \\
Co$_{20\_4}$ & $C_{1}$  & 2.39 & 2.85 & 0.98 & 1.73  \\
Co$_{20\_5}$ & $I_{h}$  & 2.44 & 1.50 & 7.85 & 1.38  \\
\end{tabular} 
\end{ruledtabular}
\end{table}

\begin{figure}
\includegraphics[width=16.5cm,angle=0]{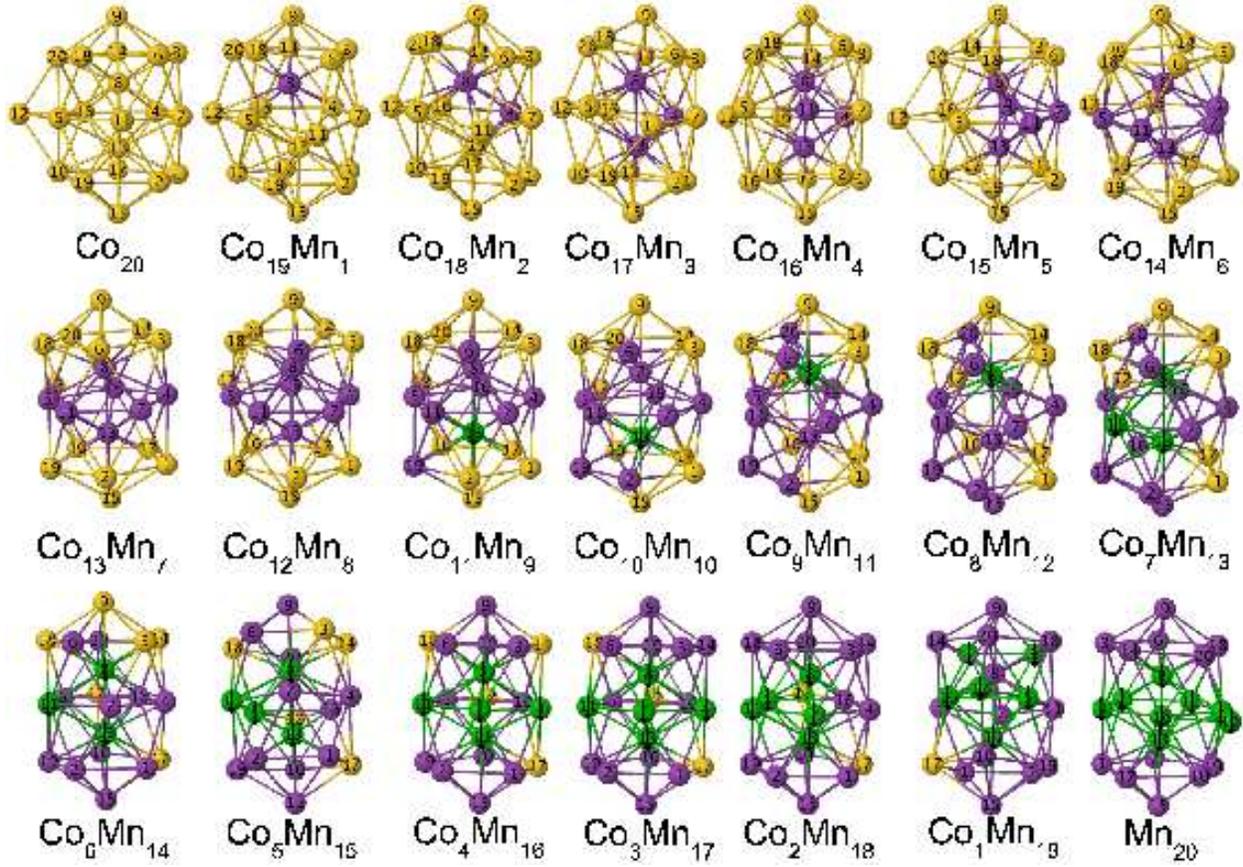}
\caption{\label{fig3}(Color online) Illustration of the ground state structures for CoMn binary clusters with
$\Omega=20$. Co atoms are shown with yellow spheres whereas Mn atoms are represented with spheres in magenta. The green
color represents the Mn atoms that couple antiferromagnetically with the rest of atoms of each cluster.}
\end{figure}

\begin{figure}
\includegraphics[width=16.5cm,angle=0]{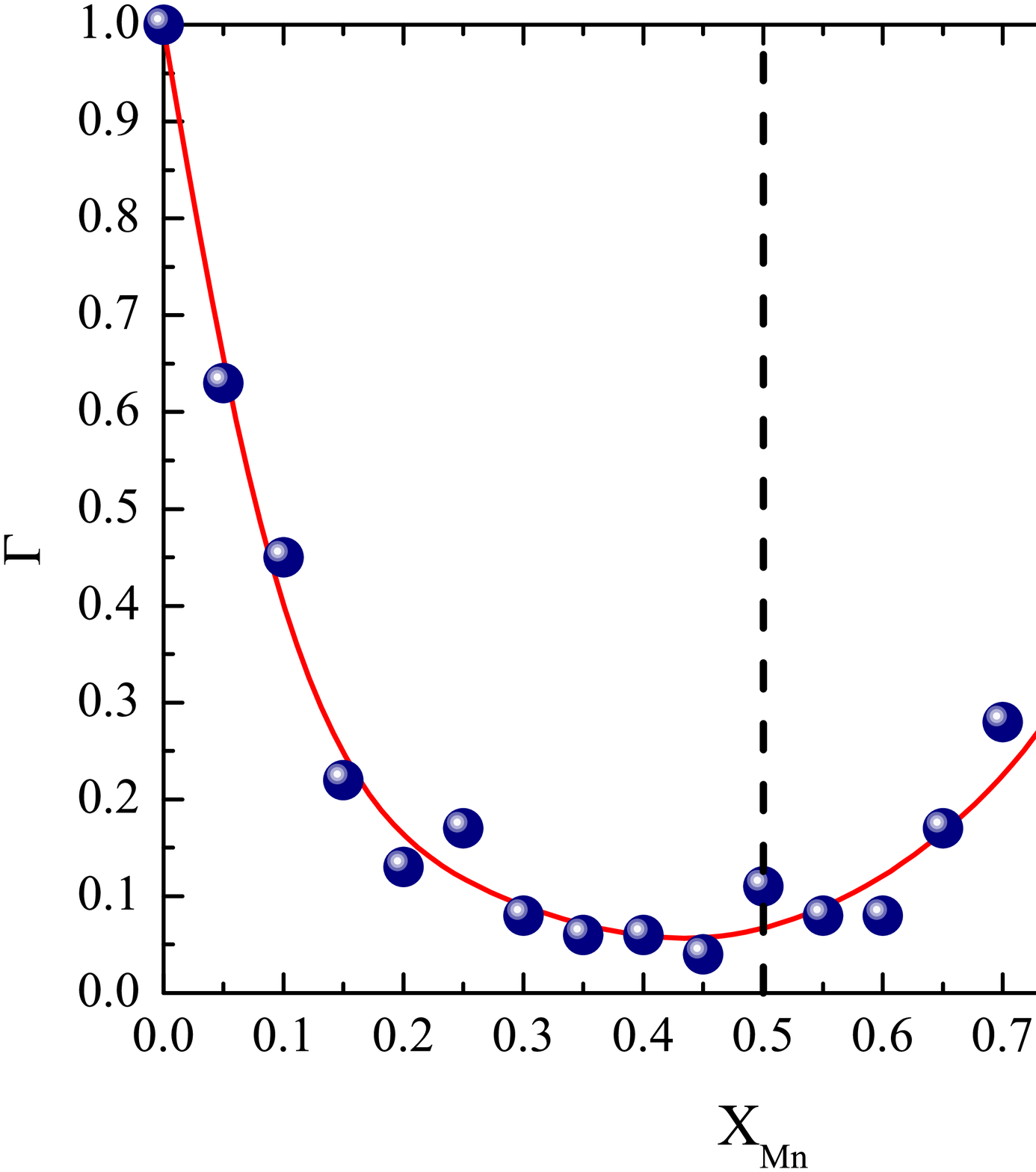}
\caption{\label{fig4}(Color online) Chemical order parameter as a function of the Mn concentration. The solid line
is a polynomial fitting to the numerical values of~$\Gamma$ and serves as a guide to the eye to appreciate the
asymmetry of the chemical order parameter with respect to the dash vertical line (midpoint of the Mn concentration).} 
\end{figure}

Another specific difficulty of alloy clusters is the great number of possibilities they offer
to distribute the solute atoms over the sites at a given geometrical arrangement of the atoms. Thus, for example, in the case of 
Co$_{10}$Mn$_{10}$ the number of possible homotops for a geometry with inequivalent positions is 
$C_{10}^{20}=\frac{20!}{10!^2}\simeq1.8\times 10^5$. With the aim of overcoming the aforementioned point, we have planned out
a new strategy consisting in optimize the geometries of the 20 homotops belonging to the Co$_{19}$Mn$_1$ cluster. As
starting point, every structure was built from 
the optimized geometry of Co$_{20}$ cluster but substituting a Co by a Mn atom at the 20 accessible positions of the cluster.
After that, we performed a geometry optimization of the 20 homotops to obtain the stability of them as a function of the
position occupied by the Mn atom. The rest of Co-Mn alloy clusters were created by replacing the cobalt atoms
by Mn atoms at the positions with the lowest energy and once again they were reoptimized in geometry. We examined the method
for the case of Co$_{18}$Mn$_2$ homotops where one of the two Mn atoms was attached to the position with the lowest-energy
and the other Mn was positioned in the rest of available sites. After the optimization procedure, 
we have observed that the lowest-energy 
structure was the one with Mn atoms in the lowest-energy positions and the energy of the other geometries was
estimated to be 3.22 eV greater than the ground-state cluster. 
We have also put the geometry optimization 
method to the test of the  experiment~\cite{koretsky} as shown in Table~\ref{table2}. The 
ionization potentials provided by our DFT
calculations are in good agreement with the available experimental results and consequently, 
our reported structures should be very 
close to the ground-state structures of the Co-Mn alloy series clusters with $\Omega=20$. The lowest-energy
structures are illustrated in Fig.~\ref{fig3} and the geometrical properties are collected in Table~\ref{table3}.
The Co-Mn alloy geometries converged in slightly distorted structures of that of the Co$_{20}$. In particular, the
distance of the atom labeled as 12 in Fig.~\ref{fig3} to the rest of atoms belonging to the cluster is gradually 
reduced with the enhancement of the Mn concentration up to the Co$_{15}$Mn$_5$ cluster which elongates the distance.
For the rest of clusters, the distance persists approximately unaltered. This point is reflected in the average nearest
neighbor distance for Co-Co bonding reported in Table~\ref{table3}. It is also important to comment that
the average nearest neighbor distances for the Co-Co, Mn-Mn and Co-Mn bondings are in general very close each other
(see Table~\ref{table3}). This results favor the assumption commented above that the addition of Mn atoms just only alters
a little bit the geometry of the ground-state structure of Co$_{20}$ cluster.
To conclude the structural discussion, and with the aim to show how the Mn atoms diffuse into the Co ones, it is convenient
to define an order parameter that is positive and close to 1 when the phase separation or segregation takes place and close
to 0 when the mixing or disorder is the main contribution to the arrangement of the Mn atoms in the cluster. The parameter
which meets the aforementioned conditions is the chemical order and it is thus defined as
\begin{equation}
  \label{eq:3}
  \Gamma=\frac{N_{\mathrm{i-i}}+N_{\mathrm{j-j}}-N_{\mathrm{i-j}}}
  {N_{\mathrm{i-i}}+N_{\mathrm{j-j}}+N_{\mathrm{i-j}}}
\end{equation}
where $N_{i-j}$ represents the number of nearest neighbor i-j bonds (with i=Co and j=Mn). The values of $\Gamma$ reported
in Table~\ref{table3} are plotted in Fig.~\ref{fig4}. It is easily observed in Fig.~\ref{fig4} the asymmetry of the
chemical order parameter with respect the dash vertical line. Thus, at low concentrations of Mn atoms $\Gamma$ decreases
more rapidly than for higher concentrations of them so that the mixing is favored in an interval of X$_{\mathrm{Mn}}$ ranging 
approximately from 0.15 to 0.65.
The reason for that behavior is attributed to the positions occupied by
the Mn atoms in the range of low concentration. As can be seen from Fig.~\ref{fig3}, the impurity atoms swap positions with
the inner Co atoms. The number of bondings between the inner positions and the rest of atoms is higher than the 
outer-shell positions. Thus, the number of Co-Mn bondings is favored for lower concentrations of the impurity and 
in consequence $\Gamma$ decrease rapidly. But however, for higher concentrations of the Mn atoms, the process 
is not inverted and the cobalt atoms occupy mainly the outer-shell positions instead of the inner ones.

\begin{table}
\caption{\label{table2}Ionization potentials (in eV) for Co${_m}$Mn${_n}$ clusters with $\Omega=20$. The
	experimental data were taken from Ref.~\onlinecite{koretsky}.}
\begin{ruledtabular}
\begin{tabular}{lcccccc}
$n$ & 0 & 1 & 2 & 3 & 4 & 5 \\
\hline
Theory & 5.45 & 5.54 & 5.60 & 5.38 & 5.51 & 5.40 \\
Experiment & 5.45 & 5.44 & 5.37 & 5.34 & 5.31 & 5.23 \\
\end{tabular} 
\end{ruledtabular}
\end{table}

\begin{table}
  \caption{\label{table3} Structural properties of the lowest-energy 
  Co-Mn alloy clusters with $\Omega=20$. We report the average nearest
  neighbor distance for the Co-Co, Mn-Mn and Co-Mn bondings. The chemical
  order defined in Eq.~(\ref{eq:3}) is also provided in the last column. 
  The interatomic distances are given in \AA.}
\begin{ruledtabular}
\begin{tabular}{lcccc}
  Cluster & d$_\mathrm{Co-Co}$ & d$_\mathrm{Mn-Mn}$ & d$_\mathrm{Co-Mn}$ & $\Gamma$    \\
\hline
Co$_{20}$        & 2.46 &      &      & 1.00   \\
Co$_{19}$Mn$_{1}$  & 2.39 &      & 2.42 & 0.63 \\
Co$_{18}$Mn$_{2}$  & 2.38 & 2.51 & 2.40 & 0.45 \\
Co$_{17}$Mn$_{3}$  & 2.38 & 2.31 & 2.40 & 0.22 \\
Co$_{16}$Mn$_{4}$  & 2.40 & 2.39 & 2.37 & 0.13 \\
Co$_{15}$Mn$_{5}$  & 2.49 & 2.43 & 2.42 & 0.17 \\
Co$_{14}$Mn$_{6}$  & 2.49 & 2.43 & 2.45 & 0.08 \\
Co$_{13}$Mn$_{7}$  & 2.50 & 2.46 & 2.43 & 0.06 \\
Co$_{12}$Mn$_{8}$  & 2.50 & 2.46 & 2.44 & 0.06 \\
Co$_{11}$Mn$_{9}$  & 2.50 & 2.45 & 2.44 & 0.04 \\
Co$_{10}$Mn$_{10}$ & 2.50 & 2.45 & 2.44 & 0.11 \\
Co$_{9}$Mn$_{11}$  & 2.50 & 2.45 & 2.44 & 0.08 \\
Co$_{8}$Mn$_{12}$  & 2.49 & 2.45 & 2.44 & 0.08 \\
Co$_{7}$Mn$_{13}$  & 2.48 & 2.46 & 2.45 & 0.17 \\
Co$_{6}$Mn$_{14}$  & 2.49 & 2.46 & 2.45 & 0.28 \\
Co$_{5}$Mn$_{15}$  & 2.48 & 2.46 & 2.46 & 0.28 \\
Co$_{4}$Mn$_{16}$  &      & 2.46 & 2.46 & 0.39 \\
Co$_{3}$Mn$_{17}$  &      & 2.46 & 2.46 & 0.55 \\
Co$_{2}$Mn$_{18}$  &      & 2.46 & 2.46 & 0.72 \\
Co$_{1}$Mn$_{19}$  &      & 2.46 & 2.47 & 0.84 \\
Mn$_{20}$        &      & 2.46 &      & 1.00 \\
\end{tabular} 
\end{ruledtabular}
\end{table}

\section{Results and Discussion}
\begin{figure}
\includegraphics[width=16.5cm,angle=0]{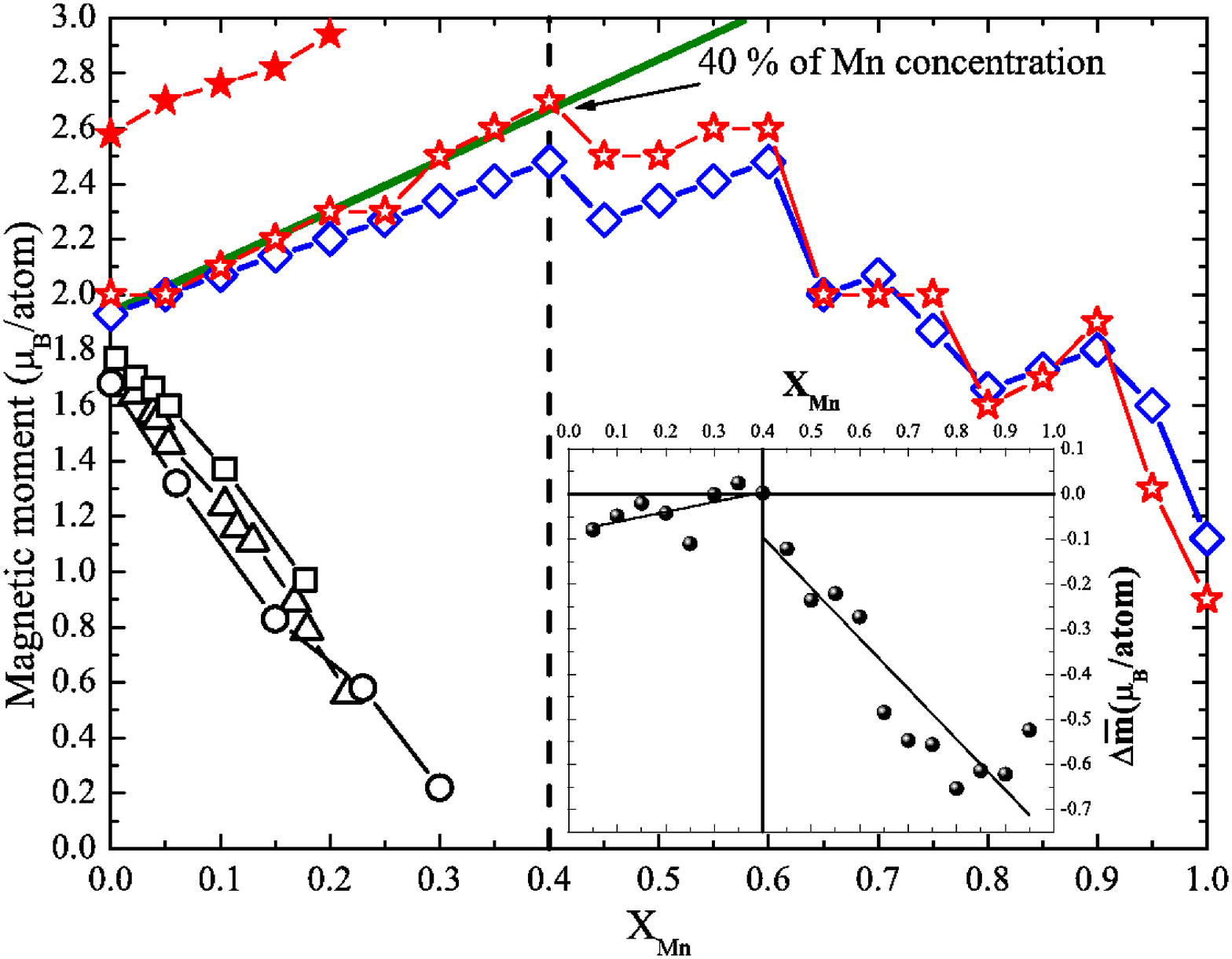}
\caption{\label{fig5}(Color online) Magnetic moments per atom of Co$_m$Mn$_n$ clusters with $\Omega=20$ calculated
in this work (empty star symbol) and measured in Ref.~\onlinecite{yin} (filled star symbol). The data for
the CoMn bulk are taken from Ref.~\onlinecite{crangle} ($\triangle$ and $\Box$ symbols) and Ref.~\onlinecite{poerschke} 
($\bigcirc$ symbol). The solid line represents the linear fitting of the magnetic moments for the clusters with 
n$\le$8 and it retains a slope of 1.83~$\mu_B$. The diamond points are the expecting magnetic moments according 
to Eq.~(\ref{eq:5}). The inset represents the difference between the magnetic moments per atom 
of the Co first neighbors of Mn atoms 
($\bar{m}_{alloy}$) in the Co$_m$Mn$_n$ and  their Co counterparts ($\bar{m}_{Co_{20}}$) in Co$_{20}$ cluster 
($\Delta\bar{m}=\bar{m}_{alloy}-\bar{m}_{Co_{20}}$) 
as a function of Mn concentration. The linear fitting is a guide to the eye.}
\end{figure}

The anomalous behavior of the SP curve in Co-Mn alloy clusters compared to the CoMn bulk \cite{crangle,poerschke} is shown in 
Fig.~\ref{fig5}.
The magnetic moments per atom (empty star symbol) increase with a slope of 1.83~$\mu_B$ up to a Mn concentration of 40\%, 
after which the average magnetic moments tend to decrease with increasing Mn concentration. The above results 
are in very good agreement with the available experimental measurements (filled star symbol) reported in Ref.~\onlinecite{yin} 
(1.7 $\mu_B$ and 40\%, respectively). 
The discrepancy with the numerical data 
(about 0.6~$\mu_B$ in average) may be related to the omission of the orbital moment contribution since the total magnetic moment
is $\langle\vec{M}\rangle=2\langle\vec{S}\rangle+\langle\vec{L}\rangle$. However, according to the results of the 
calculations performed for binary TM clusters reported in Ref.~\onlinecite{rollmann}, the orbital moments represent a 
very small correction to the total magnetic moment and consequently underpins our approximation. Likewise, the source of error 
could also be ascribed to a wrong assignment of the ground-state structures of Co-Mn alloy cluster, but however it is unlikely 
because the measured
value of the magnetic moment\cite{knickelbein} in the case of the Co$_{20}$ cluster is about 2.04 $\mu_B$ which is very close to our predicted
value of 2.00 $\mu_B$.

The negative slope 
($\sim$ 6.0~$\mu_B$ per Mn substitution) of the SP curve for the bulk CoMn alloy, which is in contrast to the cluster behavior, is explained on
the basis of the VBS concept \cite{friedel}, i. e., a resonant scattering of the conduction electrons
at the Mn sites which induces a narrow peak above the Fermi level for both spin channels. This point was confirmed by 
{\it ab initio} band
structure calculations \cite{stepanyuk}.  
In the case of the CoMn alloy clusters, the first attempt to explain the 
positive slope was also based on the assumption of the existence of a VBS but this 
time only the majority-spin VBS was conjectured
to remain below the HOMO level \cite{yin}. However, we can rule out the existence of the VBS and come to the same 
conclusion if we admit two hypothesis: first,
the assumptions of the naive RB model are valid and second, the spin-up $d$ band of the alloy is fully occupied 
\cite{ohandley}. See, as an example, Fig.~\ref{fig2}(a) above. Thus, under these conditions the alloy 
moment per average atom is given by
\begin{equation}
  \label{eq:4}
  \left<\mu_{alloy}\right>=\left<\mu_{host}\right>-c\Delta Z \mu_B
\end{equation}
where c is the impurity concentration per atom and $\Delta Z$ is the atomic number difference 
of the impurity relative to the host.
For Co host doped with Mn impurities, $\Delta Z=-2$ and the slope of the average alloy magnetic moment relative to the host is 
positive and proportional to 2 $\mu_B$, which is relatively close to the experimental value (1.7 $\mu_B$) and our reported result
(1.83 $\mu_B$). The VBS approximation can predict successfully the SP curve behavior 
in most of the TM solid alloys whereas its application to alloy clusters is less reliable because of the lack 
of a periodic crystalline lattice \cite{friedel1}. Moreover, the VBS and RB model fail to predict the behavior of the 
SP curve for Mn concentrations greater than 40 \% because they are only valid for small concentrations of impurities. 

We have derived a formula for the average magnetic moment of CoMn alloys
that corrects Eq.~(\ref{eq:4}) and takes into account the antiferromagnetic (AF) coupling between the impurities:
\begin{equation}
\label{eq:5}
\left<\mu_\mathrm{alloy}\right>=m \left<\mu_\mathrm{Co}\right>+(n-g(n)) \left<\mu_\mathrm{Mn}^\uparrow \right>-
g(n)\left<\mu_\mathrm{Mn}^\downarrow\right>
\end{equation} 
where $m$ and $n$ are the number of Co and Mn atoms in the cluster, respectively and $g(n)$ is the number of Mn atoms that 
couple antiferromagnetically with the rest of atoms \cite{footnote1} (see Table~\ref{table4}). In Fig.~\ref{fig3}, we have also plotted in green the Mn
atoms that couple antiferromagnetically to their neighbors. The averaged magnetic moments per atom and per clusters
with different Mn compositions obtained from our DFT
calculations for Co and
Mn atoms are $\left<\mu_\mathrm{Co}\right>=1.93$, 
$\left<\mu_\mathrm{Mn}^\uparrow \right>=3.30$, and $\left<\mu_\mathrm{Mn}^\downarrow\right>=2.15$ $\mu_B/atom$.     
The values (diamond points) provided by Eq.~(\ref{eq:5}) are plotted in Fig.~\ref{fig5}. To gain more insight into the physics
behind Eq.~(\ref{eq:5}) and
based on the results provided by the electronic structure calculations, we have identified a double mechanism that explains the 
magnetic enhancement trend for CoMn clusters below 40\% of Mn concentration and the successive fast dropping of the magnetic
moment. 

\begin{figure}
\includegraphics[width=16.5cm,angle=0]{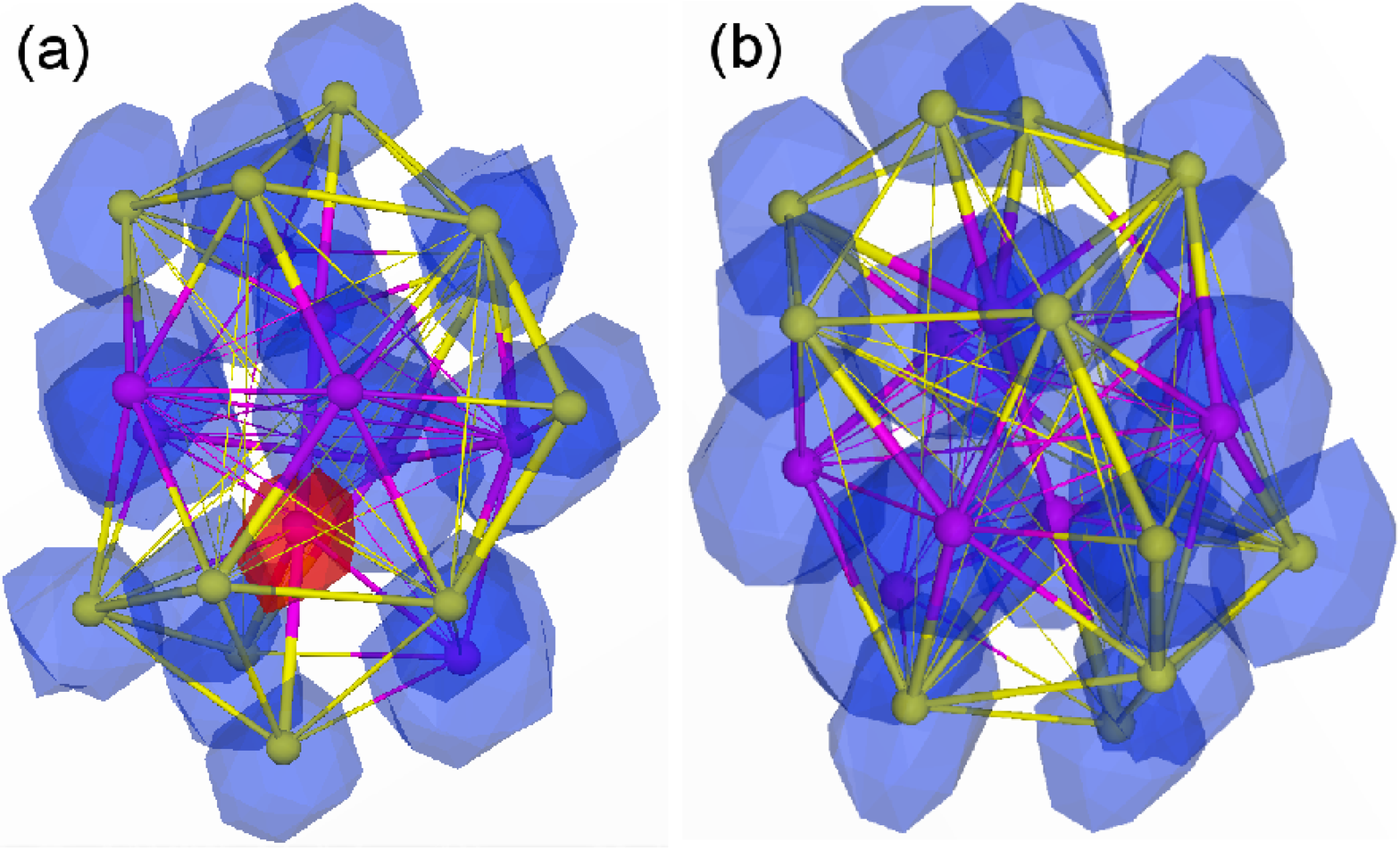}
\caption{\label{fig6}(Color online) Isosurface plot of the magnetization density ($m(\mathbf{r})=
[\rho^\uparrow(\mathbf{r})-\rho^\downarrow(\mathbf{r})]\mu_{B}$) for (a) Co$_{11}$Mn$_9$ and 
(b) Co$_{12}$Mn$_8$. The Co and Mn atoms are illustrated by the yellow and magenta spheres, respectively. 
The blue surfaces represent a positive value for the magnetization density whereas the red surface indicates
a negative value.}
\end{figure}

\begin{figure}
\includegraphics[width=16.5cm,angle=0]{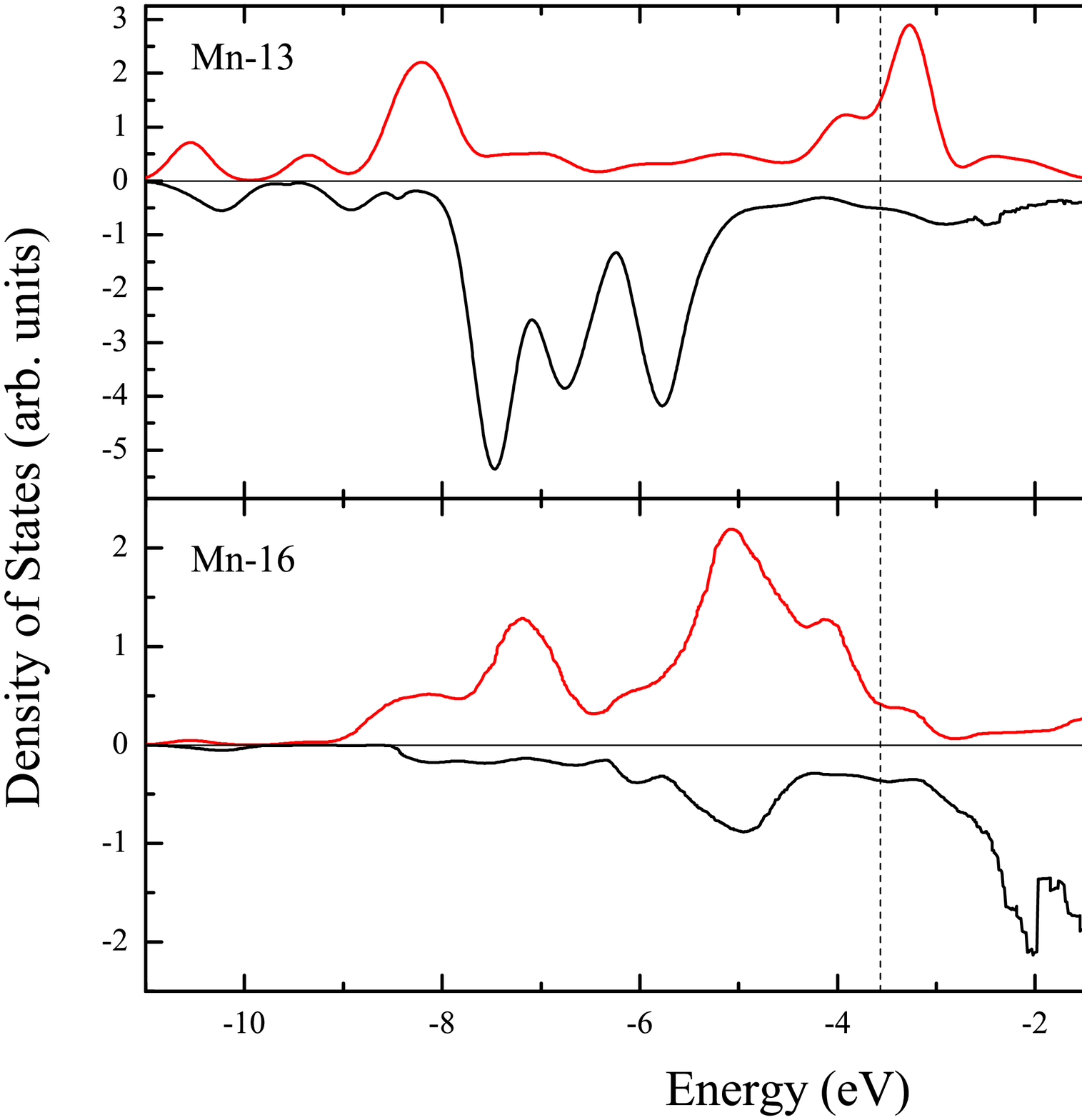}
\caption{\label{fig7}(Color online) Plot of the spin-polarized DOS for (a) Mn-13 and (b) Mn-16 atom of Co$_{11}$Mn$_9$
cluster. The upper-half area of figure (a) and (b) is for spin-up electrons while
the lower-half area is for spin-down electrons. The dotted vertical line represents the HOMO level.}
\end{figure}

The first mechanism resides in the role played by the Mn atoms, as reflected in Eq.~(\ref{eq:5}). First, the clusters increase
their magnetic moments due to the addition of the Mn moments up to a maximum of 2.7 $\mu_B/atom$  and then they decrease 
their moments because of the AF 
alignment of some Mn atoms with the rest of atoms belonging to the cluster. We have plotted in Fig.~\ref{fig6} the magnetization
density of Co$_{12}$Mn$_8$ and Co$_{11}$Mn$_9$ to see how the clusters evolve from a ferromagnetic (FM) configuration to an 
AF at a critical concentration of the impurity. The red surface surrounding the Mn atom indicates an AF coupling with the
rest of atoms while the blue surface indicates a FM alignment. The onset of Mn atoms with negative magnetic moment 
resides mainly in a Mn-Mn charge transferring. The values of N$_\mathrm{Mn-Mn}^\uparrow$ and N$_\mathrm{Mn-Mn}^\downarrow$ 
reported in Table~\ref{table4} indicate that the AF Mn atoms are the ones which establish more bondings with the rest of
Mn atoms in relation to the FM Mn atoms.
To exemplify the explanation, we have plotted in Fig.~\ref{fig7} the spin-polarized DOS for the Mn-13 and Mn-16 atoms of
the Co$_{11}$Mn$_9$ cluster. The Mn-13 atom is coupled antiferromagnetically to the rest of Mn and Co atoms 
(see Fig.~\ref{fig3}). The elevated number of bondings make that the Mn-13 atom share more electrons with its Mn environment.
This produces a charge transferring from the spin-up channel of the Mn-13 atom to the spin-up channel of the Mn atoms 
surrounding it, that is for example the case of Mn-16. Thus, at a critical number of the Mn-Mn bondings (close to 7) 
the Mn-13 atom becomes antiferromagnetic.

\begin{table}
  \caption{\label{table4} Average magnetic moments (in $\mu_B/atom$) and number of Mn atoms (g(n)) 
  that couple antiferromagnetically with
  the rest of atoms belonging to each cluster of the Co$_m$Mn$_n$ series, and with $n$ ranging from 9 up to 20. 
  The notation N$_\mathrm{Mn-Mn}^\uparrow$ 
  (N$_\mathrm{Mn-Mn}^\downarrow$) represents the average number of nearest neighbor bonds between a FM (AF) Mn atom and 
  the rest of Mn atoms.}
\begin{ruledtabular}
\begin{tabular}{lcccccccccccc}
  n & 9 & 10 & 11 & 12 & 13 & 14 & 15 & 16 & 17 & 18 & 19 & 20 \\
\hline
$\mu_\mathrm{Mn}^\downarrow$  &-0.98 & -1.11 & -1.45 & -1.50 & -2.39 & -2.49  & -2.31 & -2.73 & -2.67 & -2.27 & -2.97 & -2.95 \\
g(n)                             & 1    &  1    &  1    &  1    &  3    &  3     &  4    &  5    &  5    &  5    &  6    &  8    \\
N$_\mathrm{Mn-Mn}^\uparrow$   & 4.6 & 4.9 & 4.4 & 5.1 & 4.9 & 5.3 & 5.1 & 5.3 & 5.7 & 5.9 & 6.8 & 6.2 \\
N$_\mathrm{Mn-Mn}^\downarrow$ & 7.0 & 8.0 & 8.0 & 8.0 & 8.3 & 8.7 & 8.5 & 8.4 & 8.8 & 9.4 & 7.3 & 8.8 \\
\end{tabular} 
\end{ruledtabular}
\end{table}

The second mechanism involves also the Co atoms and the ``spin-flipping'' of the electrons belonging to the Co-Mn bonding. Although the explanation have been exemplified to the Co$_{18}$Mn$_2$ 
cluster, it is applicable to the rest of calculated clusters with $1\le n \le 8$. For these clusters, the DOS of Mn atoms
shows a higher-lying occupied bonding orbital for spin-up electrons and a lower-lying (a few eV above HOMO level) empty antibonding orbital for spin-down electrons [see
Fig.~\ref{fig2}(d)], while for Co first neighbor atoms the DOS manifests an increase of the spin-up population [see 
Fig.~\ref{fig2}(b)]. In consequence, and
based on the Mulliken population analysis, the electrons involved in the Co-Mn bonding move from the spin-down channel of Mn
atoms to the spin-up channel of the Co first neighbor atoms. This mechanism contributes to raise 
the magnetic moment of both Mn and Co 
neighbor atoms [see inset and text of the caption in Fig.~(\ref{fig5})]. However, the process stops at a 
critical concentration of the Mn atoms because
the spin-up channel of the available Co atoms does not admit more electrons. Thus, in the range with n 
varying from 9 to 19, the
process reverse and there is a 
charge transferring from the Mn spin-up channel to 
the Co neighbors spin-down one. In this case, the Mn and also the Co neighbor atoms 
[see inset in Fig.~(\ref{fig5})] decrease their moments. 

It is also interesting finally to comment that the non-monotonic decrease of the magnetic moments at the area above
40~\% of Mn concentration (see Fig.~\ref{fig5}) is attributed to the discontinuity of g(n) with the concentration
of the Mn impurity. Thus, for example, g(n) is equal to $1$ in the range $9\le n \le 12$. The average
magnetic moments of the AF Mn atoms reported in Table~\ref{table4} are approximately constant in this range and 
consequently the total magnetic moment per atom of the clusters with $9\le n \le 12$ is not reduced. The behavior changes
drastically for n=13, where g(13)=3 and $\mu_\mathrm{Mn-13}^\downarrow$=-2.39 $\mu_B/atom$. The onset of the discontinuity
in g(13) and the consequently enhancement of $\mu_\mathrm{Mn-13}^\downarrow$ causes a reduction of the total
magnetic moment of Co$_7$Mn$_{13}$ cluster.  The same kind of explanation still persists for the rest of clusters with higher
Mn concentration.

\section{Summary}
In summary, the magnetic moment enhancement of CoMn clusters below 40\% of Mn concentration and its successive fast dropping has
been explained using first-principles electronic structure calculations. The explanation resides mainly on the magnetic role played by
the Mn atoms and the ``spin-flipping'' of the electrons belonging to the Co-Mn bonding. Moreover, a new formula 
[Eq.~(\ref{eq:5})] has also been
proposed for calculating the magnetic moments of the CoMn alloy clusters. This equation represents an improvement over the 
results provided by both the RB model and the VBS approximation.

\begin{acknowledgments}
The authors acknowledge the Centro de Supercomputaci\'{o}n de Galicia for the computing facilities.
M. P. acknowledges the Isabel Barreto program for financial support. 
The work also was supported by the Ministerio de Educaci\'{o}n y Ciencia under the projet No. MAT2006-10027.
\end{acknowledgments}


\begin{thebibliography}{2}
\expandafter\ifx\csname natexlab\endcsname\relax\def\natexlab#1{#1}\fi
\expandafter\ifx\csname bibnamefont\endcsname\relax
  \def\bibnamefont#1{#1}\fi
\expandafter\ifx\csname bibfnamefont\endcsname\relax
  \def\bibfnamefont#1{#1}\fi
\expandafter\ifx\csname citenamefont\endcsname\relax
  \def\citenamefont#1{#1}\fi
\expandafter\ifx\csname url\endcsname\relax
  \def\url#1{\texttt{#1}}\fi
\expandafter\ifx\csname urlprefix\endcsname\relax\def\urlprefix{URL }\fi
\providecommand{\bibinfo}[2]{#2}
\providecommand{\eprint}[2][]{\url{#2}}

\bibitem[{\citenamefont{Bansmann et~al.}(2005)\citenamefont{Bansmann et al.}}]{bansmann}
  \bibinfo{author}{\bibfnamefont{J.} \bibnamefont{Bansmann} \bibnamefont{ {\it et~al.}}},
  \bibinfo{journal}{Surf.\ Sci.\ Rep.} \textbf{\bibinfo{volume}{56}},
  \bibinfo{pages}{189} (\bibinfo{year}{2005}).

\bibitem[{\citenamefont{Binns et~al.}(2005)\citenamefont{Binns et al.}}]{binns}
  \bibinfo{author}{\bibfnamefont{C.} \bibnamefont{Binns} \bibnamefont{ {\it et~al.}}},
  \bibinfo{journal}{J.\ Phys.\ D:Appl.\ Phys.} \textbf{\bibinfo{volume}{38}},
  \bibinfo{pages}{R357} (\bibinfo{year}{2005}).

  \bibitem[{\citenamefont{Yin et~al.}(2007)\citenamefont{Yin, Moro, and de Heer
  }}]{yin}
\bibinfo{author}{\bibfnamefont{S.} \bibnamefont{Yin}},
  \bibinfo{author}{\bibfnamefont{R.} \bibnamefont{Moro}},
  \bibinfo{author}{\bibfnamefont{X.} \bibnamefont{Xu}},
  \bibnamefont{and} \bibinfo{author}{\bibfnamefont{W.~A.} \bibnamefont{de Heer}},
  \bibinfo{journal}{Phys.\ Rev.\ Lett.} \textbf{\bibinfo{volume}{98}},
  \bibinfo{pages}{113401} (\bibinfo{year}{2007}). 

\bibitem[{\citenamefont{Rollmann et~al.}(2008)\citenamefont{Rollmann, Sahoo, Hucht and Entel
  }}]{rollmann}
\bibinfo{author}{\bibfnamefont{G.} \bibnamefont{Rollmann}},
  \bibinfo{author}{\bibfnamefont{S.} \bibnamefont{Sahoo}},
  \bibinfo{author}{\bibfnamefont{A.} \bibnamefont{Hucht}},
  \bibnamefont{and} \bibinfo{author}{\bibfnamefont{P.} \bibnamefont{Entel}},
  \bibinfo{journal}{Phys.\ Rev.\ B} \textbf{\bibinfo{volume}{78}},
  \bibinfo{pages}{134404} (\bibinfo{year}{2008}). 

\bibitem[{\citenamefont{Ganguly et~al.}(2008)\citenamefont{Ganguly, Kabir, Datta, Sanyal, and Mookerjee}}]{ganguly}
\bibinfo{author}{\bibfnamefont{S.} \bibnamefont{Ganguly}},
\bibinfo{author}{\bibfnamefont{M.} \bibnamefont{Kabir}},
\bibinfo{author}{\bibfnamefont{S.} \bibnamefont{Datta}},
\bibinfo{author}{\bibfnamefont{B.} \bibnamefont{Sanyal}},
  \bibnamefont{and} \bibinfo{author}{\bibfnamefont{A.} \bibnamefont{Mookerjee}},
  \bibinfo{journal}{Phys.\ Rev.\ B} \textbf{\bibinfo{volume}{78}},
  \bibinfo{pages}{014402} (\bibinfo{year}{2008}).

\bibitem[{\citenamefont{Shen et~al.}(2008)\citenamefont{Shen, Wang, and Zhu}}]{shen}
\bibinfo{author}{\bibfnamefont{N.} \bibnamefont{Shen}},
\bibinfo{author}{\bibfnamefont{J.} \bibnamefont{Wang}},
  \bibnamefont{and} \bibinfo{author}{\bibfnamefont{L.} \bibnamefont{Zhu}},
  \bibinfo{journal}{Chem.\ Phys.\ Lett.} \textbf{\bibinfo{volume}{467}},
  \bibinfo{pages}{114} (\bibinfo{year}{2008}).

\bibitem[{\citenamefont{Friedel}(1958)\citenamefont{Friedel
  }}]{friedel}
\bibinfo{author}{\bibfnamefont{J.} \bibnamefont{Friedel}},
  \bibinfo{journal}{Nuovo Cimento Suppl.} \textbf{\bibinfo{volume}{7}},
  \bibinfo{pages}{287} (\bibinfo{year}{1958}). 

\bibitem[{\citenamefont{Slater}(1937)\citenamefont{Slater}}]{slater}
  \bibinfo{author}{\bibfnamefont{J.~C.} \bibnamefont{Slater}},
  \bibinfo{journal}{J.\ Appl.\ Phys.} \textbf{\bibinfo{volume}{8}},
  \bibinfo{pages}{385} (\bibinfo{year}{1937});
 \bibinfo{author}{\bibfnamefont{L.} \bibnamefont{Pauling}},
  \bibinfo{journal}{Phys.\ Rev.} \textbf{\bibinfo{volume}{54}},
  \bibinfo{pages}{899} (\bibinfo{year}{1938}).

\bibitem[{\citenamefont{St-Amant et~al.}(1990)\citenamefont{St-Amant and Salahub}}]{salahub}
\bibinfo{author}{\bibfnamefont{A.} \bibnamefont{St-Amant}}
  \bibnamefont{and} \bibinfo{author}{\bibfnamefont{D.~R.} \bibnamefont{Salahub}},
  \bibinfo{journal}{Chem.\ Phys.\ Lett.} \textbf{\bibinfo{volume}{169}},
  \bibinfo{pages}{387} (\bibinfo{year}{1990}),
  \bibinfo{note}{http://www.demon-software.com/public\_html/index.html}.

\bibitem[{\citenamefont{Huzinaga et~al.}(1984)}]{huzinaga}
\bibinfo{author}{\bibfnamefont{S.}~\bibnamefont{Huzinaga}},
\bibinfo{author}{\bibfnamefont{J.}~\bibnamefont{Andzelm}},
\bibinfo{author}{\bibfnamefont{M.}~\bibnamefont{Klobukowski}},
\bibinfo{author}{\bibfnamefont{E.}~\bibnamefont{Radzio-Andzelm}},
\bibinfo{author}{\bibfnamefont{Y.}~\bibnamefont{Sakai}}, \bibnamefont{and}
\bibinfo{author}{\bibfnamefont{H.}~\bibnamefont{Tatewaki}},
 \emph{\bibinfo{title}{Gaussian Basis Sets for Molecular
  Calculations}} (\bibinfo{publisher}{Elsevier}, \bibinfo{address}{Amsterdam},
  \bibinfo{year}{1984}).

\bibitem[{\citenamefont{Pereiro et~al.}(2001)\citenamefont{Pereiro, Baldomir, Iglesias, Rosales, and Castro}}]{pereiro}
\bibinfo{author}{\bibfnamefont{M.} \bibnamefont{Pereiro}},
  \bibinfo{author}{\bibfnamefont{D.} \bibnamefont{Baldomir}},
  \bibinfo{author}{\bibfnamefont{M.} \bibnamefont{Iglesias}},
  \bibinfo{author}{\bibfnamefont{C.} \bibnamefont{Rosales}},
  \bibnamefont{and} \bibinfo{author}{\bibfnamefont{M.} \bibnamefont{Castro}},
  \bibinfo{journal}{Int.\ J.\ Quantum Chem.} \textbf{\bibinfo{volume}{81}},
  \bibinfo{pages}{422} (\bibinfo{year}{2001}).

\bibitem[{\citenamefont{Perdew et~al.}(1992)\citenamefont{Perdew, Chevary,
  Vosko, Jackson, Pederson, and Singh}}]{perdew}
\bibinfo{author}{\bibfnamefont{J.~P.} \bibnamefont{Perdew}},
  \bibinfo{author}{\bibfnamefont{J.~A.} \bibnamefont{Chevary}},
  \bibinfo{author}{\bibfnamefont{S.~H.} \bibnamefont{Vosko}},
  \bibinfo{author}{\bibfnamefont{K.~A.} \bibnamefont{Jackson}},
  \bibinfo{author}{\bibfnamefont{M.~R.} \bibnamefont{Pederson}},
  \bibnamefont{and} \bibinfo{author}{\bibfnamefont{D.~J.} \bibnamefont{Singh}},
  \bibinfo{journal}{Phys.\ Rev.\ B} \textbf{\bibinfo{volume}{46}},
  \bibinfo{pages}{6671} (\bibinfo{year}{1992}); \textbf{\bibinfo{volume}{48}},
  \bibinfo{pages}{4978} (\bibinfo{year}{1993}).

  \bibitem[{\citenamefont{Pereiro et~al.}(2007)\citenamefont{Pereiro, Baldomir, and Arias}}]{pereiro1}
\bibinfo{author}{\bibfnamefont{M.} \bibnamefont{Pereiro}},
  \bibinfo{author}{\bibfnamefont{D.} \bibnamefont{Baldomir}},
  \bibnamefont{and} \bibinfo{author}{\bibnamefont{J. E. Arias}},
  \bibinfo{journal}{Phys.\ Rev.\ A} \textbf{\bibinfo{volume}{75}},
  \bibinfo{pages}{063204} (\bibinfo{year}{2007}).

\bibitem[{\citenamefont{Wales et~al.}(1997)\citenamefont{Wales and Doye}}]{wales}
\bibinfo{author}{\bibfnamefont{D.~J.} \bibnamefont{Wales}},
  \bibnamefont{and} \bibinfo{author}{\bibfnamefont{J.~P.~K.} \bibnamefont{Doye}},
  \bibinfo{journal}{J.\ Phys.\ Chem. A} \textbf{\bibinfo{volume}{101}},
  \bibinfo{pages}{5111} (\bibinfo{year}{1997}).

  \bibitem[{foo({\natexlab{a}})}]{hyperchem}
\bibinfo{note}{HyperChem(TM) Professional 7.51, Hypercube, Inc., 1115 NW 4th Street, Gainesville, Florida 32601, USA} 

\bibitem[{\citenamefont{Fletcher}(1996)}]{fletcher}
\bibinfo{author}{\bibfnamefont{R.} \bibnamefont{Fletcher}},
  \emph{\bibinfo{title}{Practical Methods of Optimization}}
  (\bibinfo{publisher}{John Wiley \& Sons}, \bibinfo{address}{Chichester},
  \bibinfo{year}{1996}), chap.~\bibinfo{chapter}{4}, p. \bibinfo{pages}{83}.

\bibitem[{foo({\natexlab{a}})}]{database}
  \bibinfo{note}{The Cambridge Cluster Database, D. J. Wales, J. P. K. Doye, A. Dullweber, M. P. Hodges, F. Y. Naumkin F. Calvo, J. Hern\'andez-Rojas and T. F. Middleton, URL http://www-wales.ch.cam.ac.uk/CCD.html.} 

\bibitem{colby}
\bibinfo{note}{http://www.colby.edu/chemistry/PChem/scripts/ABC.html}.


\bibitem[{\citenamefont{Kittel}(2005)}]{kittel}
\bibinfo{author}{\bibfnamefont{C.} \bibnamefont{Kittel}},
  \emph{\bibinfo{title}{Introduction to Solid State Physics}}
  (\bibinfo{publisher}{John Wiley \& Sons}, \bibinfo{address}{Hoboken},
  \bibinfo{year}{2005}), p. \bibinfo{pages}{71}.

\bibitem[{\citenamefont{Koretsky et~al.}(1999)\citenamefont{Koretsky, Kerns, Nieman, Knickelbein, and Riley
  }}]{koretsky}
\bibinfo{author}{\bibfnamefont{G.~M.} \bibnamefont{Koretsky}},
  \bibinfo{author}{\bibfnamefont{K.~P.} \bibnamefont{Kerns}},
  \bibinfo{author}{\bibfnamefont{G.~C.} \bibnamefont{Nieman}},
  \bibinfo{author}{\bibfnamefont{M.~B.} \bibnamefont{Knickelbein}},
  \bibnamefont{and} \bibinfo{author}{\bibfnamefont{S.~J.} \bibnamefont{Riley}},
  \bibinfo{journal}{J.\ Phys.\ Chem.\ A} \textbf{\bibinfo{volume}{103}},
  \bibinfo{pages}{1997} (\bibinfo{year}{1999}). 

\bibitem[{\citenamefont{Crangle}(1957)\citenamefont{crangle
  }}]{crangle}
\bibinfo{author}{\bibfnamefont{J.} \bibnamefont{Crangle}},
  \bibinfo{journal}{Philos. Mag.} \textbf{\bibinfo{volume}{2}},
  \bibinfo{pages}{659} (\bibinfo{year}{1957}). 

\bibitem[{\citenamefont{Poerschke}(1991)}]{poerschke}
\bibinfo{author}{\bibfnamefont{R.} \bibnamefont{Poerschke}},
  \emph{\bibinfo{title}{Magnetic Properties of Metals: d-Elements, Alloys and Compounds}}
  (\bibinfo{publisher}{Springer-Verlag}, \bibinfo{address}{Berlin},
  \bibinfo{year}{1991}), chap.~\bibinfo{chapter}{2}, p. \bibinfo{pages}{52}.

\bibitem[{\citenamefont{Knickelbein et~al.}(2006)\citenamefont{Knickelbein
  }}]{knickelbein}
\bibinfo{author}{\bibfnamefont{M.~B.} \bibnamefont{Knickelbein}},
  \bibinfo{journal}{J.\ Chem.\ Phys.} \textbf{\bibinfo{volume}{125}},
  \bibinfo{pages}{044308} (\bibinfo{year}{2006}). 

\bibitem[{\citenamefont{Stepanyuk et~al.}(1994)\citenamefont{Stepanyuk, Zeller, Dederichs and Mertig
  }}]{stepanyuk}
\bibinfo{author}{\bibfnamefont{V.~S.} \bibnamefont{Stepanyuk}},
  \bibinfo{author}{\bibfnamefont{R.} \bibnamefont{Zeller}},
  \bibinfo{author}{\bibfnamefont{P.~H.} \bibnamefont{Dederichs}},
  \bibnamefont{and} \bibinfo{author}{\bibfnamefont{I.} \bibnamefont{Mertig}},
  \bibinfo{journal}{Phys.\ Rev.\ B} \textbf{\bibinfo{volume}{49}},
  \bibinfo{pages}{5157} (\bibinfo{year}{1994}). 

\bibitem[{\citenamefont{O'Handley}(2000)}]{ohandley}
\bibinfo{author}{\bibfnamefont{R.~C.} \bibnamefont{O'Handley}},
 \emph{\bibinfo{title}{Modern Magnetic Materials: Principles and Applications}}
  (\bibinfo{publisher}{John Wiley \& Sons}, \bibinfo{address}{New York},
  \bibinfo{year}{2000}), chap.~\bibinfo{chapter}{5}, p. \bibinfo{pages}{153}.

\bibitem[{\citenamefont{Friedel}(1954)\citenamefont{Friedel
  }}]{friedel1}
\bibinfo{author}{\bibfnamefont{J.} \bibnamefont{Friedel}},
  \bibinfo{journal}{Advances in Physics} \textbf{\bibinfo{volume}{3}},
  \bibinfo{pages}{446} (\bibinfo{year}{1954}). 

\bibitem[{foo({\natexlab{a}})}]{footnote1}
\bibinfo{note}{The numerical values of $g(n)$ provided by our DFT calculations are:

$g(n)=\left\{ 
\begin{array}{lr}
\theta(n-8.5) & 1\le n\le 12 \\ 
      \theta(n-14.5)+3 & 13\le n \le 15 \\ 
      5-\frac{(n-23)}{24}\Pi_{i=16}^{18}(n-i)  & 16\le n\le 20 
\end{array}  \right. $

where $\theta(n)$ is the Heaviside theta function.
}

\end{thebibliography}
\end{document}